\documentclass[]{mn2e}

\usepackage{amssymb}
\usepackage{graphicx}
\usepackage{amsmath}
\usepackage{graphics}

\usepackage{color}

\title[Serkowski's polarization peak] {A new interpretation of Serkowski's polarization law}

\author[R. Papoular]{R. Papoular$^{1}$\thanks{E-mail:papoular@wanadoo.fr}\\
$^{1}$Service d'Astrophysique and Service de Chimie Moleculaire,\\
CEA Saclay, 91191 Gif-s-Yvette, France}

\begin{document}

   \maketitle
\label{firstpage}

\begin{abstract}
 The basic tenets of the alternative interpretation to be presented here are that the spectral profiles of the star light polarization peaks observed in the visible and near IR are a result of the optical properties of silicate grains in the same spectral range, not of the grain size, provided it remains within the range of Rayleigh's approximation. The silicate properties are those obtained experimentally by Scott and Duley \cite{sco} for the non-iron bearing amorphous forsterite and enstatite. The whole range of observed Serkowski polarization profiles can be simulated with mixtures made of forsterite plus an increasing fraction (0 to 0.5) of enstatite as the spectral peak shifts from 0.8 to 0.3 $\mu$m. Fits to individual observed polarization spectra are also demonstrated.

The optical extinction of silicates in the vis/IR (the ``transparency range") can be understood by analogy with the thoroughly studied amorphous hydrogenated carbons and amorphous silica. It is due to structural disorder (dangling bonds and coordination defects) and impurities, which give rise to electronic states in the forbidden gap of semi-conductors. Because they are partially localized, their extinction power is dramatically reduced and has been ignored or simply described by a low, flat plateau. As their number density depends on the environment, one expects variations in the ratio of optical extinction coefficients in the visible and mid-IR.

It is also argued that the measured steep rise of extinction beyond  3 $\mu$m$^{-1}$ into the UV is due to atomic transitions, and so cannot give rise to coherent molecular polarization, but only localized  extinction.

\end{abstract}


\section{Introduction}

The polarization of star light has been, for nearly a century, the subject of extensive observation, interpretation and simulation. As a result of these, as well as astronomical emission and extinction measurements, silicates of the Olivine class emerged as the model candidate of choice, at least in the mid (MIR) and far (FIR) spectral ranges (see Mathis et al. 1977) . The ubiquitous and conspicuous 10 and 20-$\mu$m features are essential to this choice (see, for instance, the study of forsterite (Olivine group) and enstatite ( Pyroxene group) by Pegourie and Papoular 1985). In the visible, these silicates were found to be fairly transparent (see Nitsan et al.  1978, Palik 1985; Dorschner et al. 1995). This was difficult to reconcile with observations of polarization of InterStellar (IS) visible light which exhibits a wide, prominent, asymmetric bump whose peak wavelength, $\lambda_{max}$, varies from 0.3 to 0.8 $\mu$m, with a mean value of 0.55 (see Spitzer 1978; Whittet 2003, Fig. 4.10 and 4.11, Martin and Whittet 1990, Whittet 1992, Martin 1999); this range of $\lambda_{max}$ is exceptionally found to extend to about 1 $\mu$m  towards particular sources (see Goodman et al. 1995). Neither could this polarization bump be assigned to the other major IS dust component, namely carbon-rich grains, for their absorption spectrum displays no such feature.

 Half a century ago, when this bump was first observed and discussed, van de Hulst \cite{vdh} had just popularized grain absorption and scattering by grains with sizes outside the reach of Rayleigh's approximation, using Mie's theory. His calculations showed, for instance, that the extinction cross-sections (c-s) of long dielectric cylinders, having a diameter $a$, peaked around $\lambda_{max}=a$ and were slightly different in the two principal, orthogonal,  directions, for wavelengths on the red side of the peak. This ensured some polarization peaking at a wavelength $\sim 2\lambda_{max}$ \it for real  constant optical indexes, $n$,  distinctly larger than 1 and small imaginary indexes, $k$ \rm : at longer wavelengths, both c-s's decrease and, at shorter ones, their difference decreases (see Spitzer 1978,  Martin 1974). Numerical values for specific examples were tabulated by Wickramasinghe \cite{wic}. This became the basis of the theoretical model of light polarization by dust. Mathis \cite{mat83} noted that, for this to agree with polarization observations, $k$ should not exceed 0.03.

Mathis \cite{mat86} later clearly laid down the basic tenets of the model. He obtained good fits to observed polarizations (between 0.3 and 3 $\mu$m) with size distributions of infinite cylinders in $a^{-3.5}$ (the MNR distribution, see Mathis et al. 1977), with diameters $a$ between 0.005 and 0.25 $\mu$m, and \it constant complex refractive index 1.6-0.005 i, weighted by a probability of being superparamagnetic (SPM) , i.e. of containing at least one ferromagnetic domain \rm. The main fitting constraints, here, are the (MNR) size distribution law and the probability law for a grain of containing a SPM domain. The latter requires relatively large grains ($a >>100\,\AA{\ }$, Chikazumi 1964, p.240, Kittel 1986).

Later, Martin and co-workers (Kim and Martin 1994, 1995a, 1995b ), in a series of seminal papers, sought instead to deduce the best size distribution directly from polarization data, with no other constraint than the use of ``astronomical silicate'' (Draine 1985) as a model material. The population required by polarization decreases towards small $a$ values because both principal c-s's decrease with $a$, and, towards large sizes, because both tend to the same saturation value depending on $k$. \it Since observed polarization peaks straddle the visible range of wavelengths, the model imposes that the required grain sizes straddle nearly the same range of lengths, roughly 0.1 to 1 $\mu$m. \rm
 
They noted that the computed model required an extra population of very small aligned grains because of the steep rise of $k$ at the fundamental edge of the electronic energy gap of silicates,beyond wave number $\lambda>6\, \mu$m$^{-1}$ (Kim and Martin 1995a). But, on their alignment model, only larger grains are aligned. They therefore sought to avoid small grains by subsuming a different edge behavior in space than on earth.
 
I emphasize that in these studies, \it the burden of explaining the shape of this bump, is laid exclusively upon the grain size distribution, on the assumption that the 
imaginary index $k$ is very small and constant in the spectral range of observed polarization bumps.\rm 

It must be stressed that Martin et al. (see Kim et al. 1994) were fully aware that the presence of a ferromagnetic material in a grain may seriously affect polarization spectra in the visible and NIR because of absorption peaks (which were later assigned to Fe$^{2+}$). They also realized that their approach needed to be complemented by taking into account the effect of spectral variations of $k$, especially in the UV (see Kim 1995).

 The present work is based on later measurements by Scott and Duley \cite{sco} on pure, amorphous , forsterite, Mg$_{2}$SiO$_{4}$ or Fo, and enstatite, Mg$_{2}$Si$_{2}$O$_{6}$, which showed that, in the visible, their $k$'s both differ significantly from the previously assumed model silicates: while both are smaller than 0.01, they are not negligible and vary notably, regularly and characteristically, with the wavenumber. It will be shown that combinations of these profiles may mimic the polarization bump. This requires that the grain efficiency factors, $Q_{ext}$, be smaller than 1, i.e. that the grain size be small enough that Rayleigh's approximation applies as far as the visible spectral range. SPM materials are not required, and are even excluded if not in very small relative amounts, as iron has a very strong signature near 1 $\mu$m, which would be prominent in the polarized spectrum, contrary to observations (see Zeidler et al. 2011).
 
 Apart from Rayleigh's condition, this model has no constraint on grain size provided the grain is at least partially aligned with the magnetic field; high elongations (``needles") are not necessary. It also accounts for the limited range of the peak wave number of observed polarization bumps, and of the observed relation between UV/vis and IR polarization peaks.

 Section 2 describes the procedure and goals. Section 3 illustrates the indexes of refraction of the silicates used in this work, with special attention to their spectral profile in the so-called ``transparency'' range. Section 4 shows fits of the models to a number of representative Serkowski polarization curves. Section 5 shows fits to particular observed spectra.

\section{The procedure}

Consider a grain having volume $V$ and complex dielectric constant  $\epsilon=\epsilon'+\mathrm{i}\epsilon''$, with a nearly spheroidal shape. \it Assuming its size is small enough that Rayleigh's approximation may be applied, \rm the extinction cross-sections for the two principal orientations relative to the electric field of the incident light (wavelength $\lambda$) are given by

\begin{equation}
C_{i}=\frac{2\pi nV}{\lambda}\frac{\epsilon''}{(1+L_{i}(\epsilon'-1))^{2}+L_{i}^{2}\epsilon''^{2}}.
\end{equation}

where $L_{i}$ is the depolarization factor in direction $i$, $\epsilon'=n^{2}-k^{2}$ and  $\epsilon''=2nk$ ($n,k$: optical indexes; see Bohren and Huffman 1983).

Through a sight line with grain column density $N_{d}$ cm$^{-2}$ to a source of unpolarized light, the observed polarization perpendicular to the magnetic field,
 when the extinction is weak, can be approximated by

\begin{equation}
P\sim N_{d}(C_{//}-C_{\perp})/2\,,
\end{equation}

where $C_{//}$ and $C_{\perp}$  are the cross-sections of the grain when its symmetry axis is, respectively, parallel and perpendicular to the magnetic field.  It is apparent that, in between dielectric resonances, both c-s's scale like $\epsilon''/\lambda$, so  the function $PL=P(\lambda)\lambda$ scales like $\epsilon''$. But this is equal to $2nk$. As $n$ is nearly constant \it between resonances,\rm one finally gets

\begin{equation}
PL\propto k.
\end{equation}

For purposes of modeling, one may therefore compare the observed functions $PL$ with the optical indexes $k$ of candidate grain materials. 

The abundant observations of vis/NIR polarization peaks (see bibliography in, for instance, Whittet 2003) have led to their mathematical representation by the \it empirical \rm Serkowski law 

\begin{equation}
P_{\lambda}=P_{max}exp(-K\,\mathrm{ln}^{2}(\frac{\lambda_{max}}{\lambda})),
\end{equation}

where

\begin{equation}
K=c_{1}\lambda_{max}+c_{2},
\end{equation}

and $c_{1}$, $c_{2}$ are positive constants depending on the sightline (Wilking's law, 1980). This implies a systematic decrease of the feature width as  $\lambda_{max}$ increases (see Fig. 7 of Martin and Whittet 1990). Martin et al. \cite{mar99} proposed a modified Serkowski law to better fit the polarizations peaking in the near UV, but it needs more fitting parameters.

\begin{figure}
\resizebox{\hsize}{!}{\includegraphics{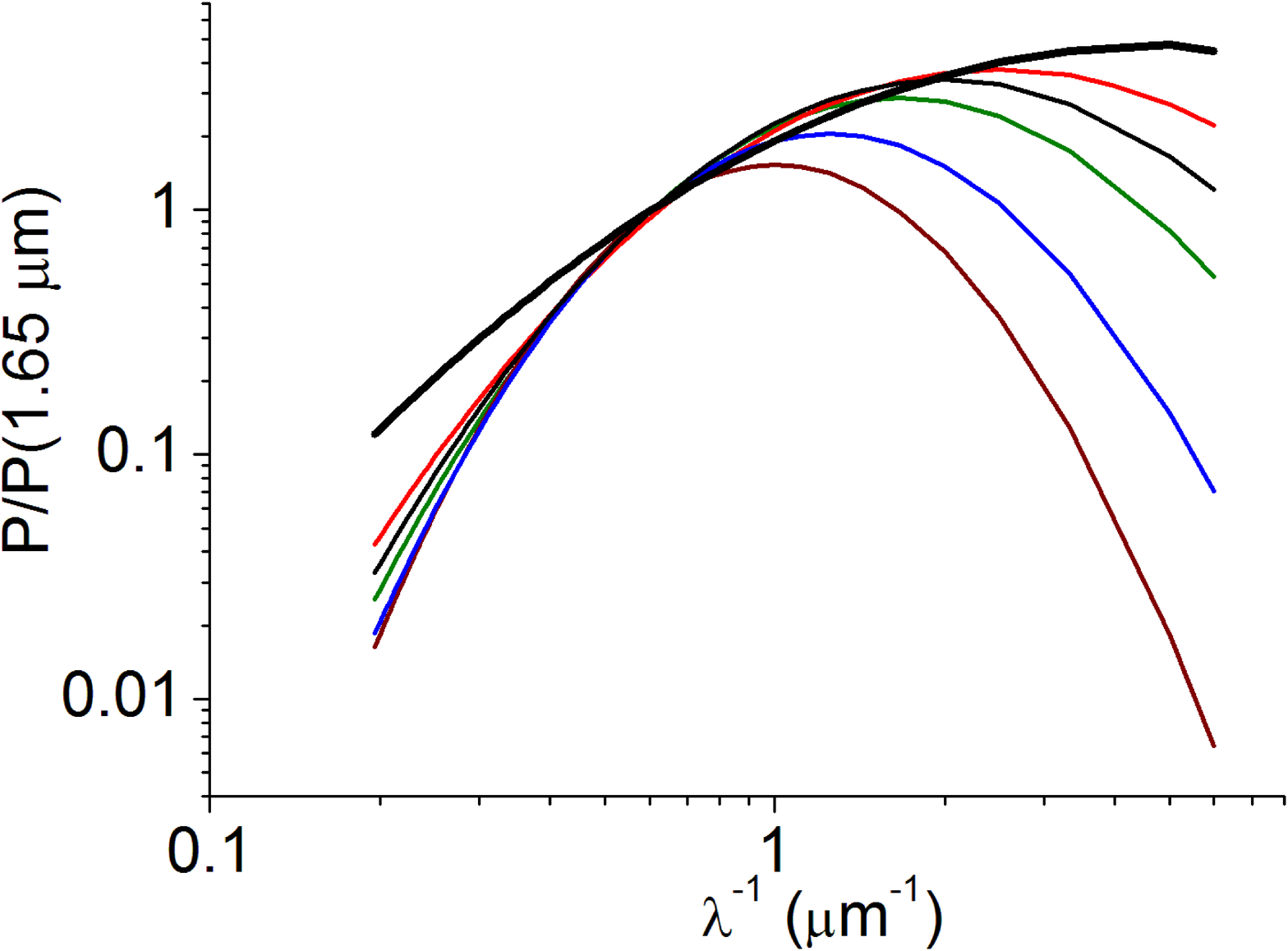}}
\caption[]{Serkowski's law for $\lambda_{max}$= 0.3, 0.4, 0.5, 0.6, 0.7 and 0.8 $\mu$m, from top to bottom, covering the range of most observed polarization spectra. In fact there is a dearth of observations beyond 2 $\mu$m$^{-1}$. All curves are normalized at 1 for $\lambda=1.65\,\mu$m as in Fig. 7 of Martin and Whittet \cite{mar90}.}
\label{Fig:serkolaw}
\end{figure}

Serkowski's law allows one to conveniently represent the wealth of observational measurements, keeping in mind that it does not  necessarily fit individual source measurements exactly, especially not outside the visible range for which it was parametrized: in the IR, for instance,  a power law seems to be more appropriate (see Martin 1999). Since the range of observed $\lambda_{max}$ is quite limited, this can be done by drawing Eq. 1 for a small number of values of the latter. Following in the steps of Martin and Whittet \cite{mar90}, Fig. \ref{Fig:serkolaw} displays the result for $\lambda_{max}=$0.3, 0.4, 0.5, 0.6, 0.7 and 0.8 $\mu$m, using $c_{1}=1.7, c_{2}=0.01$, as proposed by Whittet \cite{whi03} for the vis/NIR range. 

 Given  that extinction polarization is proportional to the small difference between light extinctions for two orthogonal alignments of a non-spherical grain, one expects a close similarity between the profiles of extinction and polarization, both for the slopes and the features. While this is indeed generally true in the MIR, no obvious feature corresponding to the visible polarization bump has been detected in the ISEC (InterStellar Extinction Curve); but, in this range, the extinction by graphitic grains is already dominant and smooth. The extinction which gives rise to this bump must therefore be weak and probably due to the other major dust component, namely silicates. Inversely, the ISEC  features a conspicuous peak at 2175 \AA{\ }, generally assigned to the $\pi$   electronic resonance of small graphite chunks (see Papoular et al 2013), which is rarely apparent in the polarization observations. How this might come about is discussed below.

\section{The optical indexes of amorphous forsterite and enstatite  measured by Scott and Duley}

Natural Olivines from Earth, planets and meteorites have been extensively analyzed by means of a wide variety of laboratory and \it in situ \rm techniques (see, for instance Clark 1999, Simmons and Tilley 2008, MINDAT data base 2018, Dyar et al. 2009). Their generic elementary composition is of the form X$_{n}$SiO$_{4}$, where X$_{n}$ stands for Mg$_{2}$, Fe$_{2}$, Mn$_{2}$, CaMg, CaFe, CaMn. 

The optical properties of these materials were found to adequately mimic the 2 characteristic, lattice vibration, features near 10 (stretching) and 20 $\mu$m (bending) observed in the sky (see, for instance,  Pegourie and Papoular 1985,  Beichman et al. (edrs) 1985 IRAS Explanatory Suppl., JPL Pasadena;  v Dishoeck 2004, and references therein).

In the vis/NIR range, iron-bearing silicates like fayalite Fe$_{2}$SiO$_{4}$ are very absorbing  because  of a strong and relatively narrow feature at 1 $\mu$m, due to un-coordinated Fe$^{2+}$ . This makes them unfit for modeling the Serkowski peaks. By contrast, silicon dioxide SiO$_{2}$ (glass), as well as MgO are known to be relatively transparent between 0.1 and 10 $\mu$m ($ibid.$). Their absorption index has consequently been generally considered as negligible in this so-called transparency range, or taken to be weak and nearly flat, which is no better for present purposes. While this may be partially true for perfect crystalline forms, it is much less so for amorphous states. 

\begin{figure}
\resizebox{\hsize}{!}{\includegraphics{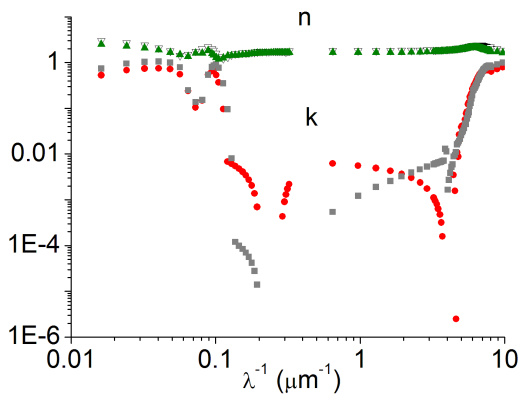}}
\caption[]{Optical indexes of the amorphous silicates measured by Scott and Duley \cite{sco}. Forsterite: olive triangles for $n$, red dots for $k$; Enstatite: inverted black triangles for $n$ and gray squares for $k$. Both $n$'s are small, flat and close enough that the extinction cross-sections closely scale like $k/\lambda$ according to Eq. 3.}
\label{Fig:nk}
\end{figure}

More sensitive and accurate measurements of the optical properties of earthly and planetary silicates are now available than were a few decades ago. They show that pure SiO$_{2}$, MgO and Mg$_{2}$SiO$_{4}$ (\it forsterite \rm) are nearly transparent in the visible, much more so than the ``astronomical silicates'' defined by Draine \cite{dr85} to represent IS dust (e.g. Zeidler et al. 2011, Pitman et al. 2013). However, Scott and Duley \cite{sco} deduced the complex index of refraction of amorphous \it forsterite \rm and \it enstatite \rm  (MgSiO$_{3}$ or Mg$_{2}$Si$_{6}$O$_{6}$) from reflection measurements in the range 0.1 to 20 $\mu$m. They found that, in the transparency range of these two non-iron bearing silicates, there remains some weaker extinction with different trends in strength as a function of wavelength. This is evidenced by illustrating their data in a log-log graph, as in Fig. \ref{Fig:nk}.
 
Scott and Duley obtained sample films by excimer laser ablation of geological samples of polycrystalline forsterite (Fo) and enstatite (En). They checked that the elemental ratios of parent materials were preserved in the measured films. 

 Absorption is indeed found to be quite weak, but not negligible, in the transparency range. More importantly it is not flat and its spectral profile is not the same for the two silicates because the ideal coordinations are different, and perhaps the disorder too. In the case of forsterite, it has the shape of a bump peaking in the visible, while for enstatite it rises steadily towards the blue. This suggests that mixtures of the two materials in different proportions should display a tendency for the bump to widen as it shifts to the blue. A first step in this direction was taken by Papoular \cite{pap18}, who showed that polarization by pure, amorphous forsterite grain sizes of order 0.05 $\mu$m provide a decent fit to the Serkowski peak of the source HD283701, at least in spectral profile. Here, it will be shown that mixtures of amorphous   forsterite and enstatite are able to fit reasonably well the whole range of observed bumps. 

The laboratory data cannot be used without some understanding of the physics behind the $n,k$ curves. A detailed study of amorphous hydrogenated \it silicon \rm was provided by Street \cite{str}, and explains the general physical concepts of a semi-conductor and of the disorder which characterizes amorphous states. A similar study of graphite, amorphous carbon and amorphous hydrogenated carbons (a-C/H) had previously been made by Robertson \cite{rob}. Both are essential for the present purposes, as are the more recent and seminal books by Kuzmany \cite{kuz} and Egerton \cite{ege}.

 A semi-conductor is characterized by its valence and conduction electronic energy bands separated by a band gap. In the valence band, the outer atomic electrons remain bound to their respective nuclei, while, in the conduction band, they acquire some mobility under the force of an electric field. To force a transition from the valence to the conduction band, an incident photon must have an energy higher than the band gap energy. In a crystal, bands are separated by steep edges.
The main changes, with an amorphous semi-conductor, relative to the perfect crystalline state, are 

a) The band edges limiting the valence and conduction electronic bands become less steep because of tails extending into the forbidden gap from below and above, respectively. In silicates, the gap extends from $\sim0.1$ to $\sim\,5\mu$m.

b) Coordination defects necessarily appear, in which each atom adopts a co-ordination state different from that of most others and from the crystalline ``ideal'' (including dangling bonds and vacancies). By definition, these defects are spatially localized and their wave functions hardly spread over one another in the system. The effect of disorders of various types are vividly illustrated by Robertson \cite{rob} in the case of a-C:H. He shows that, while there is a deep gap (transparency range) in the density of (electronic) states (DOS) of diamond, the introduction of disorder (mixture of sp$^{2}$ and sp$^{3}$ bonds) fills it with $\pi$ states, yielding some extinction, whose intensity depends on the relative abundances of the two bonds.

c) In the conduction band, electron mobility is reduced more or less by disorder. In other words,  the coherence between waves scattered by individual atoms is more or less destroyed.

In a metal, there is no band gap and electrons are not bound to the constitutive atoms. They can be treated as in a plasma, where they respond coherently to the ambient electric field.

These notions can be applied to silicates: although they are insulators, not semi-conductors, they also have a \it transparency \rm gap between two adjacent, highly absorbing regions below 0.2 and above 5 $\mu$m (Fig. \ref{Fig:nk} ). Beyond 5 $\mu$m (i.e. below 0.2 $\mu$m$^{-1}$), a silicate behaves like a dielectric: the electrons are bound to the nuclei and can be treated using the Lorentz paradigm for the dielectric functions and the Rayleigh approximation for the absorption and scattering cross-sections. These imply that the individual scatterers (the atoms) are polarizable and that the waves they scatter in response to the incident wave bear some coherence with each others, so they can, in turn, give rise to a coherent wave; otherwise, the simplifying assumption of a uniform field in the grain volume (the electrostatic approximation) cannot be used for the treatment of small spheroidal grains. This treatment is also applicable to the case of a metal, because of the coherence of the response of its free electrons, which confers polarizability to the associated plasma (see Bohren and Huffman 1983). The spectrum of coherent scattering in silicates includes mainly lattice vibrations: the two, stretching and bending, fingerprints and the phonons which cover a much wider range and give rise to the apparent continuum if the host sample is large enough. Their extension to the far IR is limited by the size of the grain, and, towards the vis/UV, by the spatial atomic periodicity (see Papoular 2014 ). For usual IS molecules, the phonon vacuum wavelength is at least a few micrometers. The band intensity of phonons in disordered systems is usually 10 to 100 times weaker than that of fingerprints.

Consider now the UV spectral range. An energetic incident photon may excite an electron into the conduction band (an ``optical transition''). In a perfectly ordered structure, conduction electrons are in ``extended'' (or ``delocalized'', or Bloch) quantum states where they can transit from one atom to a neighbor; their mobility is quite high, being limited only by weak crystal fields. In an imperfect medium, electron propagation is hampered more or less by structural and compositional defects. By contrast, \it in disordered silicates (which are insulators),\rm  there is no conduction band at all. Some of the photo-excited electrons  return to their ground state by emitting a photon of the same energy; these electrons may provide some local coherence, and only in the forward direction, at that. The others excite phonons and, so, heat the grain and contribute to the continuum. As a result, the trajectory of the incident wave can only be traced by constructing its wave front, like in geometric (ray) optics, or, more generally, by having recourse to Stokes' parameters $S_{1-4}$ (see van de Hulst 1957, who clearly rationalizes the differences between the various treatments of electric response and their domains of validity). Similarly, in Quantum Mechanics, bound states are distinguished from scattering states. In the first case, Perturbation Methods seek to find the changes in discrete energy levels  and eigenfunctions of a system when a small disturbance is applied; polarizability is a consequence of such changes. In the second case, transitions between states are involved and Scattering Matrix methods compute their probability; the notion of polarizability is not invoked.

In the absence of bulk polarizability, the ``uniform field approximation'' cannot be used, nor can the continuity conditions at the borders. \it The grain's extinction cross-section reduces to the product of the number of scatterers along the sight line and their excitation cross-section (oscillator strength). The incident light cannot be polarized by the grain. \rm  It must be stressed that the terms \it coherent \rm and \it incoherent \rm used here apply only to the light-matter interaction phenomena considered in this work. One may also use the terms \it collective \rm or \it correlated \rm , and \it isolated \rm or \it localized \rm response

The theoretical understanding of light scattering in the band gap of amorphous insulators is less extensive. Of course, ``narrow'' features have been identified and assigned, such as the 1-$\mu$m feature of iron ions (see Zeidler 2011). But many sensitive laboratory measurements of optical spectra of amorphous solids devoid of heteroatoms (impurities) reveal a continuum that is still weaker than the phonon continuum and extends to the fundamental edge, near 0.1 $\mu$m, at $k$ levels that may reach $10^{-3}-10^{-2}$ (see Dorschner et al. and bibliography therein, 1995), which has not yet been interpreted. One source of this continuum may be harmonics and overtones of the above mentioned phonons. Another could be of \it electronic \rm .origin, as suggested in item (b) above. Sources of vibrational (lattice) origin are also known, such as localized phonons associated with localized electrons and defects or with non linear response (see Weaire and Taylor 1980, \it Vibrational properties of amorphous solids \rm). They are designated as polarons, solitons, breathers, etc.. The randomness of the host structure makes for a high density of such events and, hence, some coupling between them, which could add up to a collective response. The extreme weakness of the extinction continuum could be due to weak coupling between neighboring localized sources and/or non-linear response to the incident light, and/or the weakness of electronic polarizability in general

Again, the ambiguity between coherent and incoherent response arises from the fact that a measurable $k$ does not imply a collective response defined by a polarizability , such as the Lorentz-Lorenz response. The use of the latter to interpret transmission measurements already implies that a collective response is assumed \it a prioiri \rm. Until more sophisticated laboratory techniques are put to use (such as Raman spectroscopy, polarization and ellipsometry), one is left to make conjectures. 
Here, Scott and Duley's data will be used down to 0.25 $\mu$m, assuming that, in our silicates, coherence persists down to 0.25 $\mu$m, which covers the spectrum of impurities/defects in the band gap. 

\begin{figure}
\resizebox{\hsize}{!}{\includegraphics{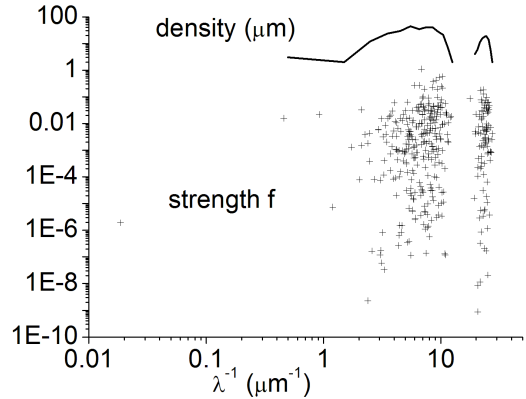}}
\caption[]{Oscillator strength (crosses) and density of oscillators (line) as a function of wave number, of an elementary cell of forsterite, Mg$_{2}$SiO$_{4}$
. These electronic oscillators contribute to IS extinction but not polarization, for they do not oscillate coherently as do the lower-frequency, molecular, oscillators.}
\label{Fig:UVforst}
\end{figure}

On the other hand, the spectrum of incoherent scattering is given by the ensemble of excitations of atomic electrons to upper electronic states of the host system.  Figure \ref{Fig:UVforst} shows this spectrum for an elementary cell of forsterite, Mg$_{2}$SiO$_{4}$, as computed using the HyperChem v8.0 modeling software package, and the incorporated semi-empirical (mixed classical/quantal) computational method PM3, described in detail by the manufacturer, Hypercube, in their publication, HC50-00-03-00. The computation procedure is utterly different than the computation of lattice vibrations. It delivers the probability of transitions from the highest occupied to the lowest unoccupied molecular orbitals. Plotted in the figure are the  oscillator strengths and their density for transition energies as high as 100 eV, as a function of wave number. In a solid, of course, the orbitals merge into bands. However, the figure conveys a rough idea of the spectrum of $intra-atomic$ electronic excitation in the amorphous powder. The domain of incoherent scattering is seen to extend slightly into the visible. For bulk matter in this domain, $n$ is close to 1 while $k$ may be large, but these quantities cannot be used in Eq. 1 to compute a scattering or absorption c-s, as there is no polarization by the grain, as explained above. Accordingly, the $n,k$ data below 0.25 $\mu$m will not be used here. However, polarization is observed down to 0.125 $\mu$m (see Sec. 5), which suggests that a wall-like border between coherent and incoherent spectral regions is not physically sound and that the coherent domain, in fact, extends slightly below 0.25 $\mu$m. Therefore, a tentative extrapolation, based on Serkowski's law, will be considered for $k$, as explained in the next section.

Incidentally, this discussion can be extended to the case of the interstellar 2175-\AA{\ } feature, which is usually assigned to graphite grains but raises a similar problem: it dominates the ISEC but does not show up in polarization spectra. This will be taken up in the Appendix.

\section{Simulating Serkowski peaks with mixtures of pure, amorphous forsterite and enstatite}

\begin{figure}
\resizebox{\hsize}{!}{\includegraphics{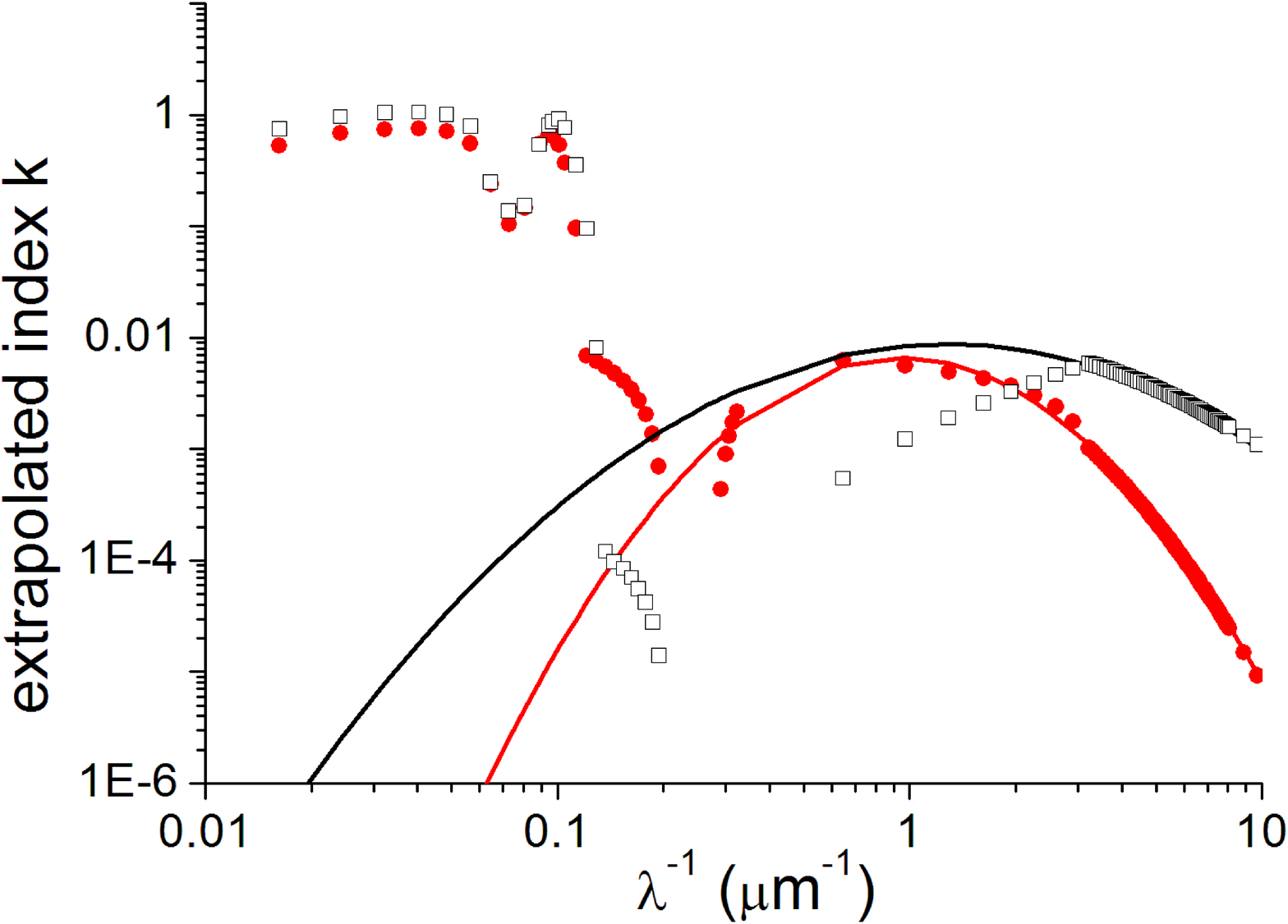}}
\caption[]{The 2 sets of $k$ data to be mixed for simulation purposes: forsterite: red dots; enstatite: open squares. The solid lines are the corresponding Serkowski's profiles for $\lambda_{max}=0.7$ and 0.3 $\mu$m, respectively, that were used to extrapolate beyond 3 $\mu$m$^{-1}$.}
\label{Fig:kcorr}
\end{figure}

As stated in the previous section, the measured $k$'s in the UV include coherent and incoherent processes. The latter were excluded by removing data beyond  3 $\mu$m$^{-1}$. They were substituted with a tentative extrapolation of the ``coherent'' $k_{fo}$ and $k_{en}$ into the UV range, inferred from Serkowski's law and relation (3) for $\lambda_{max}=0.7$ and 0.3 $\mu$m, respectively. This choice is inspired by the observation that the UV spectra of the extreme Serkowski cases are best simulated by pure, amorphous forsterite and  maximum-enstatite content, respectively. The 2 sets of $k$ data, selected and extrapolated, that were mixed for simulation purposes are plotted in Fig. \ref{Fig:kcorr}, together with the corresponding $PL$ functions for $\lambda_{max}=0.7$ and 0.3 $\mu$m.

Note that no similar extrapolation was attempted for the trough in $k$ between 0.2 and 0.3 $\mu$m$^{-1}$ (the valence edge). In this range, Martin and Whittet  \cite{mar90} and Martin et al. \cite{mar92} observed that Serkowski's law did not fit observations of particular sources. These were better fitted with power laws with an index $\beta$ ranging roughly between 1.6 and 2.5. While a drop in $k$ is certain to be present, laboratory measurements on various silicates indicate that the trough may be partly filled by cosmic ``impurities'' such as OH, Fe, Cr, etc. (see Zeidler et al  2011 and references therein). This subject deserves further investigation.

\begin{figure}
\resizebox{\hsize}{!}{\includegraphics{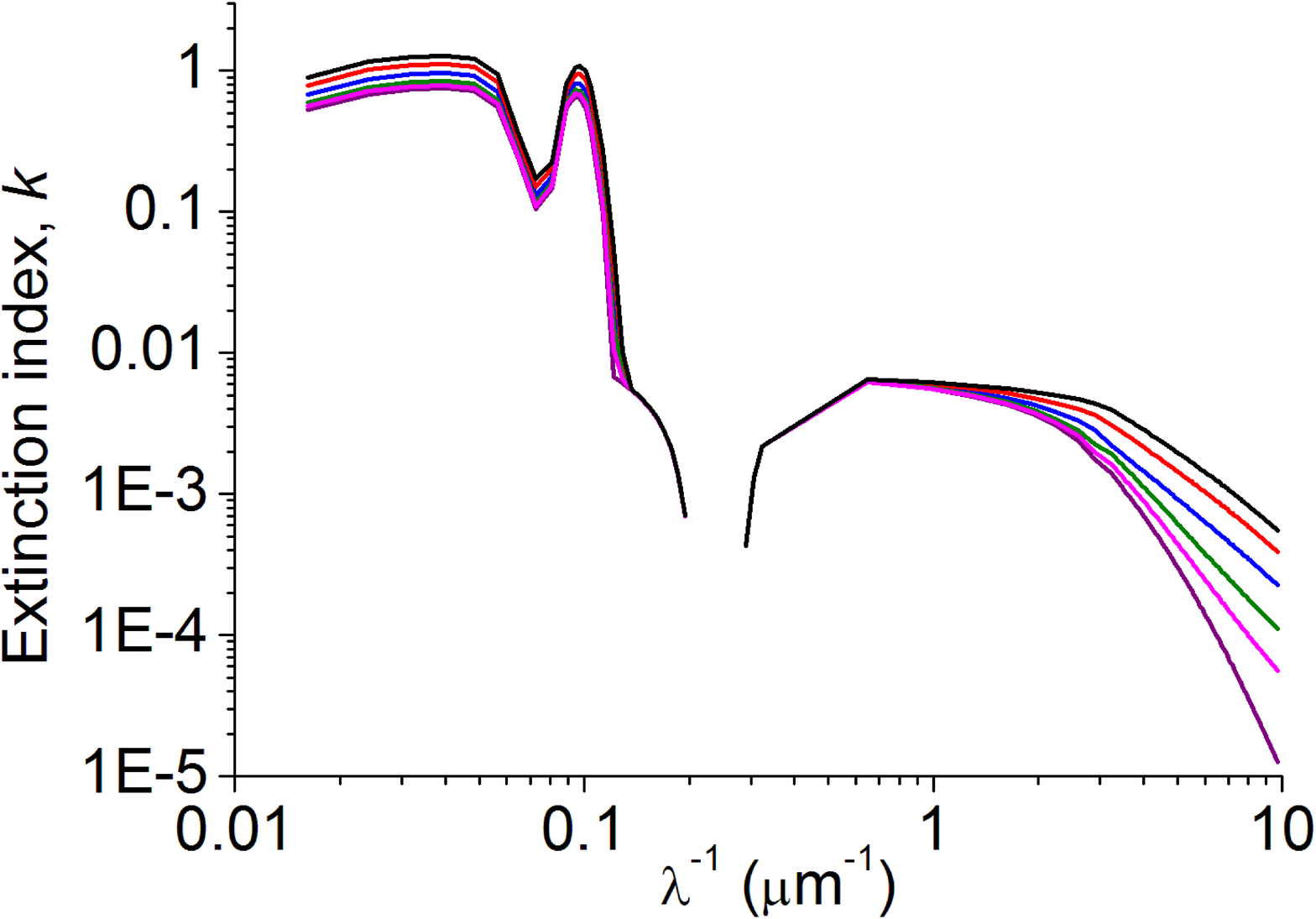}}
\caption[]{From bottom up, the 6 sets of $k$ data obtained, for $\lambda_{max}=0.8, 0.7, 0.6, 0.5, 0.4, 0.3 \mu$m, by mixing the 2 sets of Fig. \ref{Fig:kcorr} in increasing fractions of enstatite, as plotted in Fig. \ref{Fig:r}.}
\label{Fig:kmix}
\end{figure}

\begin{figure}
\resizebox{\hsize}{!}{\includegraphics{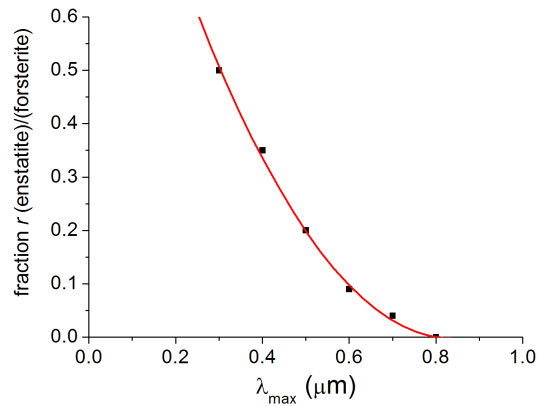}}
\caption[]{The mass ratio, $r$, of enstatite to forsterite in the model mixture, as a function of $\lambda_{max}$.}
\label{Fig:r}
\end{figure}

Finally, Fig. \ref{Fig:kmix} shows 6 sets of $k$ data obtained by mixing the 2 sets of Fig. \ref{Fig:kcorr} in increasing fractions of enstatite, as plotted in Fig. \ref{Fig:r}.

The fitting procedure naturally requires a specific choice of grain size and shape. An oblate spheroid was assumed with principal dimensions $a=b=27.5$ \AA{\ } and $c=5$ \AA{\ }. The reduction factor, $RF$, which characterizes the imperfection of alignment was found to be about 0.7 (Papoular 2018). The free parameters to be tailored were the column density of grains and the fraction, $r$, of enstatite that was added to a constant amount of forsterite: $k=k_{fo}+r\times k_{en}$. 

\begin{figure*}
\resizebox{\hsize}{!}{\includegraphics{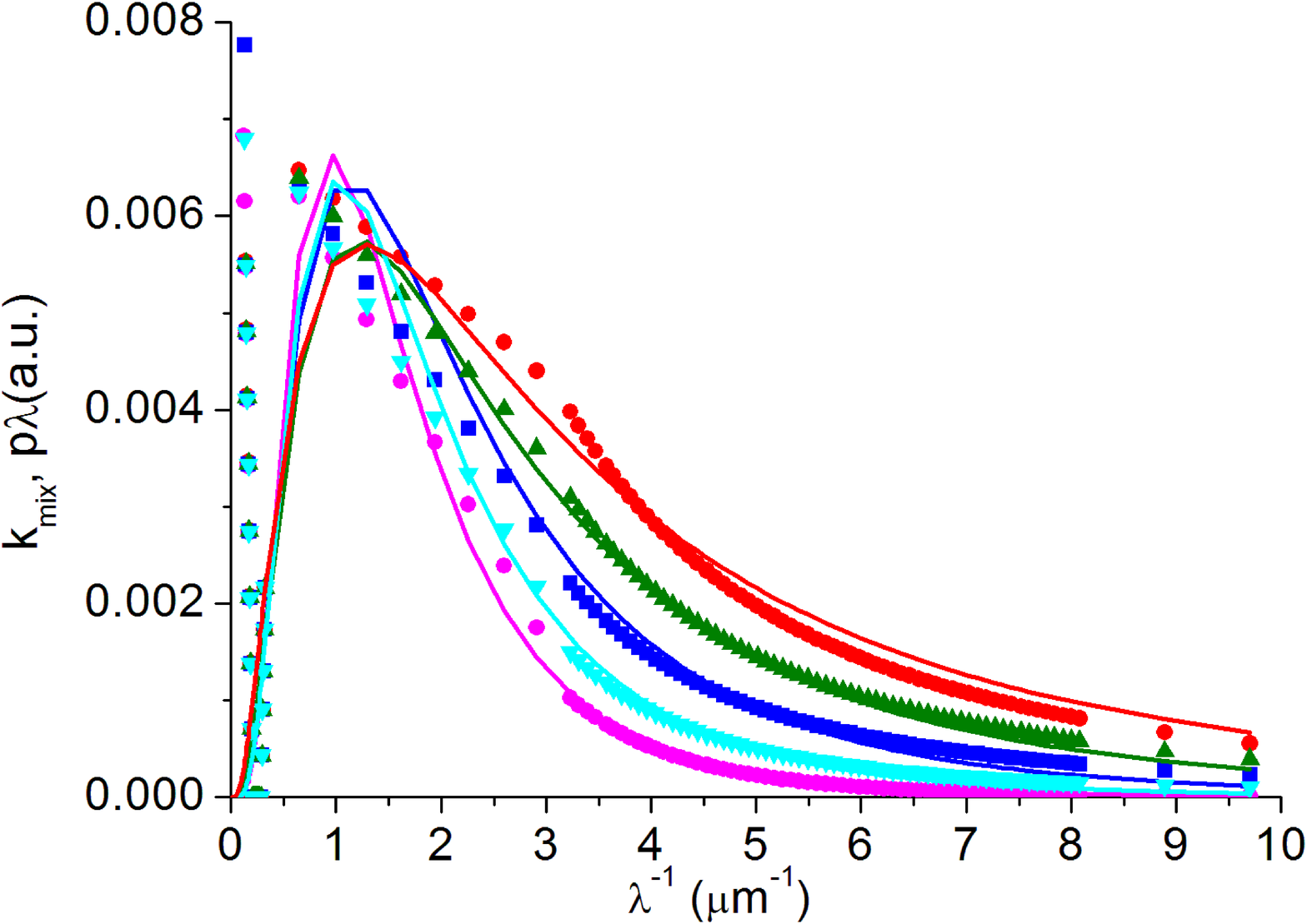}}
\caption[]{Simulations of typical Serkowski polarization peaks (lines) with mixtures of pure, amorphous forsterite and enstatite (symbols) as tabulated by Scott and Duley \cite{sco}  below $\lambda^{-1}=3 \mu$m$^{-1}$, (Fig. \ref{Fig:nk}), and extrapolated beyond that as explained in the text. In fact, what is plotted in each case is the product of the normalized empirical Serkowski law (Fig. \ref{Fig:serkolaw}) into the wavelength, and the sum $k$(Forsterite)+$rk$(Enstatite), where $r$ is the tailored parameter. The  the ratio, $r$, of enstatite to forsterite masses is 0, 0.08, 0.2, 0.35, 0.5 in order of decreasing $\lambda_{max}$: 0.7, 0.06 0.05, 0.04, 0.03 $\mu$m and increasing width (bottom up; corresponding colors: magenta, cyan, blue, olive, red). This is intended as a guide for fitting each particular observed spectrum.}
\label{Fig:fits}
\end{figure*}

Based again on Eq. 3, we now try to fit functions $PL$ with mixtures of forsterite and enstatite in various proportions. Five such fits are illustrated in Fig. \ref{Fig:fits}. The solid lines represent Serkowski's law for $\lambda_{max}=0.3, 0.4, 0.5, 0.6, 0.7 \mu$m. The symbols represent corresponding models obtained with decreasing enstatite fractions, $r$. The case of $\lambda_{max}=0.8 \mu$m was excluded to avoid confusion.

While the fits are only approximate, the disparities relative to the corresponding Serkowski laws are not inordinate in view of observed departures of accurate polarization measurements from the Serkowski formula (see, for instance, Martin et al. 1999; here, Sec. 5, Fig. \ref{Fig:HD204}). Figure \ref{Fig:fits} thus vindicates the basic tenets of the model: \it the profile of the polarization spectrum (peak wavenumber and width) derives from that of the optical indexes of the adopted silicates, not from a grain size distribution. \rm Several comments are in order.

1) The figure clearly shows a trend of decreasing $\lambda_{max}$ as the fraction of enstatite increases. This is in line with the positive slope of enstatite's $k$ in Fig. \ref{Fig:nk}, which helps lifting up the polarizations at short wavelengths. Figure \ref{Fig:r} illustrates this trend. 

2) Reasonable  simulations for $\lambda_{max}$ shorter than 0.3 $\mu$m and longer than 0.8 $\mu$m were not possible. The range in between these is remarkably close to the range of most observed values of $\lambda_{max}$, which confirms the coherence of the present model.

3) The observed polarization spectra are all smooth curves with downward concavity. If large amounts of iron-bearing silicates were  present in the aligned grains, they would show up as a prominent, relatively narrow peak near 1 $\mu$m: this is borne out by Zeidler's measurements of the absorption index of the San Carlos olivine, Mg$_{1.96}$Fe$_{0.16}$Si$_{0.89}$O$_{4}$, which displayed a feature peaking at $k(1\, \mu$m)=$4\,10^{-4}$, i.e. about one tenth of that of forsterite, with less than half its width. 
This appears to set a limit of at most a 10 \% fraction of iron relative to magnesium. Note that the cosmic abundance of iron is bout 1/4 that of magnesium.

4) Since both the polarization peak in the visible and the pair of peaks in the mid IR are assigned to the same silicates, the ratio of their intensities is determined by the corresponding ratios of extinction indexes through Eq. 3 (with some interference from the refraction index, because of the grain eccentricity). With the optical indexes used here, it should therefore remain around 0.1. However, a compendium of observations gathered by Martin and Whittet \cite{mar90} (in their Fig. 4 and 5), seems to  suggest that it may reach about 2 towards several sources. On the other hand, the \it Planck \rm satellite recently measured polarizations through the diffuse IS medium, in the visible and the sub-millimeter ranges;  Papoular \cite{pap18} (and references therein) tried to fit these data using ``astronomical silicate'' (Draine 1985).for the grain properties. It appears, from this attempt, that the ratio of the peak polarizations in the visible and at 10 $\mu$m is about 0.5. One should therefore consider the possibility that some silicates in space may be still more disordered than those of Scott and Duley, and/or harbor much more defects in the electronic gap, thus significantly enhancing $k$ in the visible (the bump). A factor 10  would bring the peak visible extinction of this model slightly above the level of the transparency plateau of ``astronomical silicate''. Other evidence to the same effect is given in Sec. 5.

5) Observations show that the ratios $P_{V}/A_{V}$ and  $P_{V}/\tau_{V}$ for a given $A_{V}$ or $E_{B-V}$ are limited upwards at about 3\%, with a great dispersion below this limit. Assuming the column densities of silicates and carbons are, respectively, $N_{sil}$ and $N_{c}$, and the corresponding cross-sections are $C_{sil}$ and $C_{c}$, then 

\begin{equation}
 P_{V}/A_{V}=\frac{0.5 f_{pol}RF N_{sil}\Delta(C_{sil})}{ N_{sil}C_{sil}+N_{c}C_{c}}\,,
\end{equation}

 where $f_{pol}$ is the fraction of polarized silicate grains, RF the reduction factor, $\Delta C$ is the cross-section difference between the 2 principal directions, and all cross-sections are for the visible wavenumber, 0.547 $\mu$m. With a grain eccentricity $a/c\sim0.2$,  $\Delta C/C=0.48$, so we must have 
\begin{equation}
\frac{f_{pol}}{1+N_{c}C_{c}/N_{sil}C_{sil}}\leq1/16.
\end{equation}
If, for instance, the contribution of silicates to extinction is 1/10 th of that of carbon dust in the visible, then, the fraction of polarizing silicate dust should be $\sim$1/2 or less. However, other factors may come into play, such as the orientation of the magnetic field relative to the observer, or collisions with ambient gas atoms.

6) The polarizing grain size adopted here is small enough that scattering remains negligible with respect to absorption in the spectral range of interest. 

\begin{figure}
\resizebox{\hsize}{!}{\includegraphics{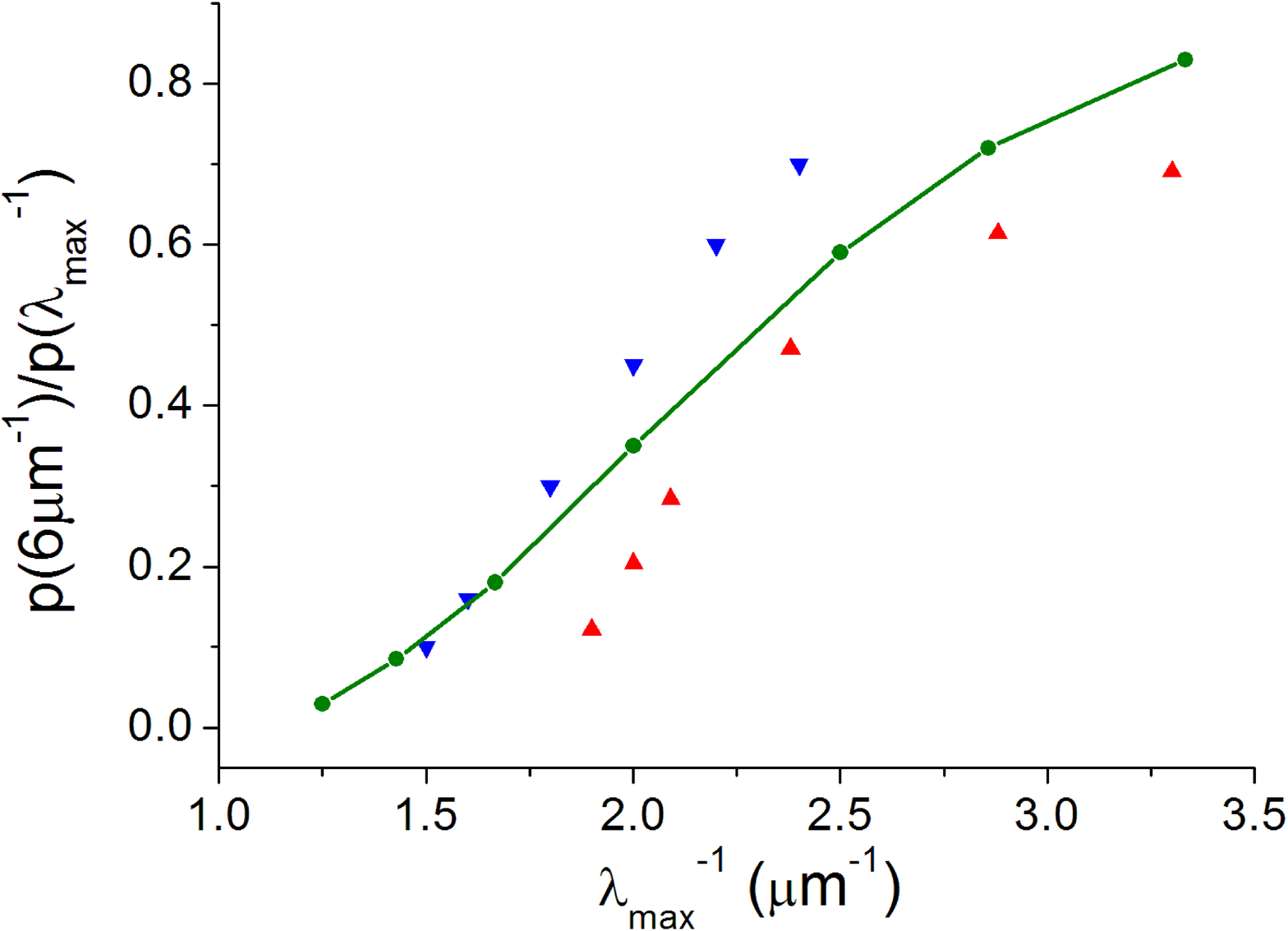}}
\caption[]{The spectral profile of polarization in the UV. \it Red triangles: \rm  present computations using Scott and Duley's optical properties of pure, amorphous  forsterite and enstatite. \it Olive dots and line \rm: Serkowski formula. \it Inverted blue triangles \rm: adapted from Martin et al. \cite{mar99}. See discussion in text.}
\label{Fig:p6}
\end{figure}

7) Martin et al. \cite{mar99} studied in great detail IS  polarization in the UV, where it often deviates considerably from Serkowski's formula. An adapted form of their results is plotted in Fig. \ref{Fig:p6}, as the ratio of measured polarizations at 6 $\mu$m$^{-1}$ and at the peak. The same marker is also plotted as deduced from Serkowski's formula and from the present simulations. The 3 curves have roughly similar shapes, reflecting the same physical trends, but are clearly shifted from one another. This highlights the above-mentioned deviations, as well as possible deviations of Scott and Duley's silicates from those in IS space. The maximum peak wavenumber shift between the present simulations and the Serkowski curves to which they were fitted is about 0.5 $\mu$m$^{-1}$

8) Observations reported by  Whittet \cite{whi03} (Fig. 4.12; see references therein) suggest some correlation between $\lambda_{max}$ and the normalized selective IS extinction, $R_{V}$, in the form

\begin{equation}
R_{V}=(5.6\pm0.3)\lambda_{max}.
\end{equation}

In the present model, this correlation cannot be  directly related to the aligned silicate grains because a) their extinction in the vis/IR is too weak relative to that of the carbonaceous grains, and b) $E_{B-V}$ is negative (see Fig. \ref{Fig:kmix}). If $R_{V}$ increases, it must be because $A_{B}$ decreases relative to $A_{V}$. A likely cause of this is a reduced incidence of the 2175 \AA{\ } feature with its steep negative slope. This, in turn, implies that the graphitized component of carbonaceous IS dust decreases relative to its amorphous component (see Papoular et al  1995). On the other hand, if $\lambda_{max}$ increases concomitantly according to Eq. 8, this must be due to the decrease of the amount of enstatite,  MgSiO$_{3}$  relative to forsterite, Mg$_{2}$SiO$_{4}$, in the polarizing dust, according to Fig. \ref{Fig:r}. Both trends indicate an increased oxygenation of the environment. This might indirectly correlate the trends of $R_{V}$ and  $\lambda_{max}$.

\section{Particular cases}

The present model will now be confronted with observations of particular sources.

\begin{figure}
\resizebox{\hsize}{!}{\includegraphics{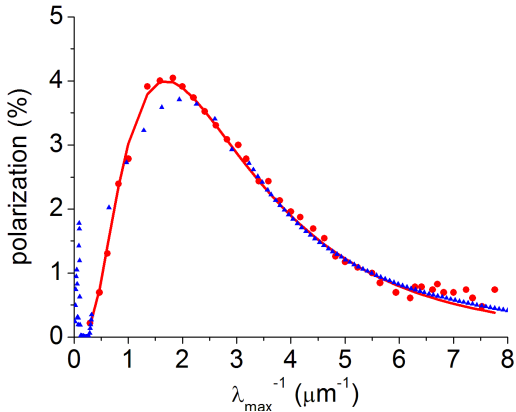}}
\caption[]{Fit of pure, amorphous forsterite (\it blue triangles \rm) to the observed polarization spectrum of HD 161056 (\t red dots \rm, adapted from Somerville et al. 1994) and its corresponding Serkowski curve for $\lambda_{max}=0.59\,\mu$m$^{-1}$ (red line). In this case, no enstatite is required.}
\label{Fig:HD161}
\end{figure}

One of the most extended available UV/vis/NIR spectrum, compiled from various observational data, particularly those 
delivered by the \it Hubble Space Telescope \rm, looking at the source HD161056, and  published by Somerville et al. \cite{som}, is plotted in Fig. \ref{Fig:HD161}, together with the corresponding Serkowski fit. The model data are drawn as blue triangles. The discrepancies concentrate near the peak and do not exceed 10\%. As expected, the fit is best in the UV tail where the $k$ extrapolation (3 to 8 $\mu$m$^{-1}$) is derived from the Serkowski formula, as explained in Fig. 4. The fitting data were computed using the grain dimensions defined above, with a reduction factor $RF=0.7$ and a grain column density $8\,10^{15}$ cm$^{-2}$. If the $k$'s in the bump are multiplied by 10, as suggested  in comment 4 of Sec. 4, then this density is reduced to $8\,10^{14}$. 

The observations clearly deviate from the Serkowski formula from 6 to 8 $\mu$m$^{-1}$. Still larger and spectrally more extended deviations have been observed, such as toward HD7252 ($ibid$). Such deviations may help better characterizing the UV properties of the polarizing dust.

\begin{figure}
\resizebox{\hsize}{!}{\includegraphics{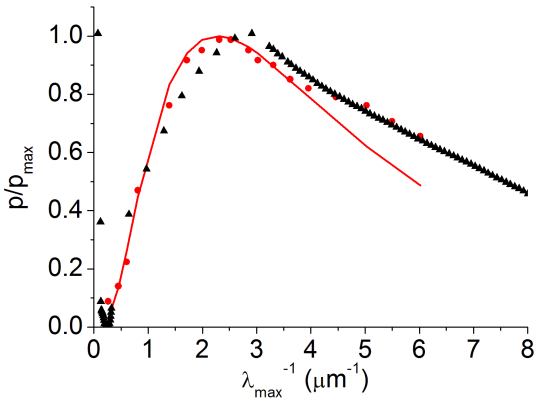}}
\caption[]{Fit of a mixture of pure, amorphous forsterite and enstatite (\it black triangles \rm) to the observed polarization spectrum of HD 204827 (\t red dots \rm, adapted from Martin et al. 1999) and its corresponding Serkowski curve for $\lambda_{max}=0.44\,\mu$m$^{-1}$ (red line). Note the increasing deviation of measurements from the Serkowski curve, upward of 3 $\mu$m$^{-1}$. This cannot be mistaken for the 2175-\AA{\ } feature. In this case, a fraction $r=0.35$ of enstatite is required to account for this behavior.}
\label{Fig:HD204}
\end{figure}

Keeping in mind that the Serkowski law is only empirical, and tailored mainly to the vicinity of the peak polarizations, the fit must be refined for each particular sight line. Consider, for instance the source HD204827, whose polarization data points (adapted from Martin et al. 1999) are drawn in Fig. \ref{Fig:HD204}. The red line is a plot of Serkowski's formula with $\lambda_{max}=0.44\,\mu$m, to fit the peak region. Obviously, this does not fit the higher wave numbers; nor can the 2175-\AA{\ } feature
 be invoked, for the discrepancy extends over a much wider range  than the width of that feature. By contrast, a mixture of forsterite with 35 \% of enstatite accounts for the UV range.
 
\section{Conclusion}
The main results of this work are as follows.

1) An alternative is proposed to the current interpretation of the polarization bump observed in the NUV/vis/NIR IS starlight. It holds that this bump follows the spectral profile of the extinction index of specific silicates in the ``transparency '' region (gap), and is not linked to the special behavior of the extinction cross-sections of grains just outside the range of validity of Rayleigh's approximation (beginning of $Q_{ext}$ saturation).

2) Thanks to the difference in spectral profiles of the extinction indexes of the measured pure, amorphous silicates, the observed correlated variations of the bump's width and its peak wave number can be reasonably simulated by mixing an increasing fraction of enstatite to a given amount of forsterite, as the peak wave number increases. The $k$'s used here are those measured by Scott and Duley \cite{sco}.

3) The occurrence of (weak) extinction in the transparency range is interpreted as due to structural and/or elemental defects in the amorphous samples. The expected  variations of these with the ambient medium in space may explain the observed variations in bump intensity relative to the intensity of the vibrational features at 10 and 20 $\mu$m, as well as discrepancies between the present model and observations.

4) The frequent occurrence of irregular excess deviations of starlight polarization from the Serkowski formula beyond 3-4 $\mu$m$^{-1}$ is interpreted as due to the grain electric response  being mostly ``incoherent'' in this range: it is due to energetic photons exciting valence electrons into uncorrelated upper quantum states (unlike metallic electrons). These cannot be treated by the  (uniform) electrostatic field approximation, generally used to compute extinction cross-sections of Rayleigh grains. In this case, there is extinction but no polarization. However, one cannot exclude the existence, in the same spectral range, of sparsely distributed ``coherent'' (vibration-like) states amenable to (weak) polarization. Hence the observed small excess polarization deviations in precisely the same spectral range. Laboratory experiments are called for to help lift ambiguities between extinctions associated with polarization and those that are not.

5) The same notion of incoherent (non-collective) electronic response explains the absence of a polarized 2175-\AA{\ } feature even in the presence of strong extinction at the same wavelength.

6)  The observation of a positive correlation between the normalized selective extinction, $R_{V}$ with the polarization peak wavelength is tentatively explained by a common sensitivity to the ambient oxygen content.

\section{Acknowledgments}

I am grateful to Prof. W. Duley for precious information regarding his experiments and comments on a draft of the paper, and to Prof. Th. Posch for giving me access to the Jena data bank, and for useful insights.

 \section{Appendix: Why is the 2175-\AA{\ } feature absent in polarization spectra although it is so strong in absorption?}
 
 The 2175-\AA{\ } absorption feature appears conspicuously at about 4.6 $\mu$m$^{-1}$ with FWHM $\sim1\,\mu$m$^{-1}$. If it were present in polarization spectra, it should emerge distinctly. Instead, what is observed is irregular excess polarization over and above the Serkowski UV tail, extending widely from 3-4 to 8 $\mu$m$^{-1}$ and varying in shape and intensity from source to source (e.g. HD204827, Fig. \ref{Fig:HD204}). 
 
 Bulk ordered graphite (e.g. HOPG, Highly Oriented Polycrystalline Graphite) displays a peak in $\epsilon''$ at $\sim3.4\,\mu$m$^{-1}$ due to electronic transitions from the $\pi$ valence band to the $\pi$* conduction band, where they are relatively free to travel along the graphene planes. The 2175-\AA{\ } feature has been assigned to the associated surface resonance (or plasmon) in small grains satisfying Rayleigh's criterion, for an incident electromagnetic wave having its electric vector perpendicular to the constitutive graphene planes (see, for instance,  Draine and Lee 1984, Papoular et al. 2013). Such a feature arises whenever $\epsilon'=-2$. In space, ``graphite'' grains cannot be expected to consist of a single stack of extended graphene layers like ideal, bulk graphite. They are more likely to be randomly oriented BSU's (Basic Structural Units) about 10 $\AA{\ }$ in size each, composed of a few small parallel graphene layers, either free-flying or assembled in larger grains, possibly interspersed with amorphous carbon of some sort. The electrostatic approximation might be applied to each BSU, but not to the whole grain. Consequently, there is no collective response of the grain, only isolated, localized, uncorrelated responses. Hence, no polarization, only extinction. Here, the BSU's play the same part as do individual atoms or molecules in silicates (Sec. 3 above).

Another effect may contribute to counteract collective effects: except in so-called ``free-electrons'' metals, one cannot ignore single electron (or localized) excitation (see Egerton 2011, Chap. 3) for, even though the plasmon is a collective phenomenon, its energy at any given moment is likely to be carried by only one electron (Ferrell 1957), which can interact with one or a few neighboring atoms and thus deposit its energy in the form of heat. Indeed, in the laboratory, the plasmon life time is observed to be very short. This indicates that some of the available oscillator strength is subtracted from the collective response, thus reducing the polarization power of the grain.

More quantitatively, it was suggested that a criterion for neglecting collective effects altogether could be obtained by stating that the number of valence electrons available for  the formation of a plasmon is small, which translates into Im($-1/\epsilon'')\ll\pi$ (Egerton, ibid, p. 134). Thus, while this quantity (the Loss Function) reaches 30 at its peak, for Aluminum, the peak is found to be as low as 1.5 for graphite (Papoular et al. 2013). 

Silicates raise a similar paradox: $\epsilon''$ exhibits a conspicuously steep absorption edge from 4 to 10 $\mu$m$^{-1}$. This is expected to translate into a distinct jump in polarization of star light by silicate grains. No such discontinuity is observed in fact. Why? According to Ichikawa \cite{ich}, electronic excitations in this spectral range may be classified as
 
 a) single-electron intra-atomic/molecular transition, as in Fig. \ref{Fig:UVforst};

b) excitation of a single electron from its ground state (here, in the valence band, such as $\pi$) to one of the upper crystal bands, such as $\pi$* or $\sigma$*; 

c) collective excitation (plasmon).

The distinction between these is clear in the case of graphite thanks to extensive studies. In the case of amorphous silicates, $\epsilon'$ is never even negative; hence, silicate grains cannot carry surface plasmons. The rise of $\epsilon''$ beyond 4 $\mu$m$^{-1}$ indicates the occurrence of electronic losses of types a) and/or b). The latter is excluded as such since the band theory does not apply; the remaining option is a).The effect of the dense environment is reduced to coupling between nearest neighbors. This affects the energies of the individual lower and upper orbitals defining the electronic transitions (a); the UV spectrum of (bulk or powder) forsterite should therefore differ from that of Fig. \ref{Fig:UVforst}, but only  in the details.

 \end{document}